**Dispersion Relation for MHD Waves in Homogeneous Plasma**

V.S. Pandey and B. N. Dwivedi
*Department of Applied Physics, Institute of Technology*
*Banaras Hindu University, Varanasi-221 005, India*
E-mail: pandey_vs@yahoo.com

**Abstract.** We consider viscosity and thermal conductivity as dissipation mechanisms to derive a general dispersion relation for MHD waves propagating in a homogeneous plasma. We show that the actual dispersion relation for MHD waves in a homogeneous medium must be six-order. The finding is in agreement (except some coefficients) with the results of Porter et al. (1994) but it is in disagreement with the previous results obtained by Kumar et al. (2006). We also discuss in detail differences between our approach and those considered by other authors.

## 1. Introduction

Coronal heating by magnetohydrodynamic (MHD) waves has been investigated extensively, beginning with Braginskii (1965) and followed by several authors (see, e.g., Zweibel 1980; Habbal and Leer 1982; Gordon and Hollweg 1983; Cargill and Hood 1989; Porter et al.1994; Laing and Edwin 1995, Pekünlü et al. (2001); Kumar et al. 2006). In this paper we derive the dispersion relation showing intermediate results and discuss differences between our approach and those considered by other authors.

## 2. General dispersion relation

We consider viscosity and thermal conductivity as dissipation mechanisms. Equations of conservation of mass, momentum, magnetic flux, energy and the equation of state are :

$$\frac{\partial \rho}{\partial t} + \nabla \cdot (\rho \mathbf{v}) = 0, \tag{1}$$

$$\rho \frac{D\mathbf{v}}{Dt} = -\nabla p + \frac{(\nabla \times \mathbf{B}) \times \mathbf{B}}{4\pi} - \nabla \cdot \Pi, \tag{2}$$

$$\frac{D\mathbf{B}}{Dt} = \nabla \times (\mathbf{v} \times \mathbf{B}), \tag{3}$$

$$\frac{Dp}{Dt} + \gamma p (\nabla \cdot \mathbf{v}) = (\gamma - 1)(Q_{th} + Q_{visc} - Q_{rad}), \tag{4}$$

and

$$p = 2nk_B T = \frac{2\rho}{m_p} k_B T, \tag{5}$$

where $\rho$, $n$, $\mathbf{v}$, p, $\mathbf{B}$, $\gamma$ and T respectively are the total mass density, electron number density, velocity, total pressure, magnetic field vector, ratio of specific heats and temperature. $\Pi$ is viscosity tensor, $Q_{th} = \nabla \cdot \kappa \nabla T$, where $\kappa$ is thermal conductivity tensor, $k_B$ is Boltzmann constant, $m_p$ is proton mass, $Q_{visc}$ is rate of viscous heating per unit volume, and $Q_{rad}$ is rate of radiative loss per unit volume. We take $Q_{visc} = -\pi_{\alpha\beta}(\partial v_\alpha / \partial x_\beta)$ from Braginskii (1965) and $Q_{rad} = n_e n_H \lambda(T)$ from Bray et al. (1991). Following Porter et al. (1994) we consider uniform background magnetic field, $B_0$, directed along the z-axis and homogeneous background plasma, with constant equilibrium values $\rho_0, T_0, p_0$ and $\mathbf{v}_0 = 0$. We linearize Eqs. (1) – (5) under the first – order approximation and obtain:

$$\frac{\partial \rho_1}{\partial t} + \rho_0 (\nabla \cdot \mathbf{v}_1) = 0, \tag{6}$$

$$\rho_0 \frac{\partial \mathbf{v}_1}{\partial t} = -\nabla p_1 + \frac{(\nabla \times \mathbf{B}_1) \times \mathbf{B}_0}{4\pi} - \nabla \cdot \mathbf{\Pi}, \tag{7}$$

$$\frac{\partial \mathbf{B}_1}{\partial t} = \nabla \times (\mathbf{v}_1 \times \mathbf{B}_0), \tag{8}$$

$$\frac{Dp_1}{Dt} - \frac{\gamma p_0}{\rho_0} \frac{D\rho_1}{Dt} = (\gamma - 1)(Q_{th} + Q_{vis} - Q_{rad}), \tag{9}$$

and

$$\frac{p_1}{p_0} = \frac{\rho_1}{\rho_0} + \frac{T_1}{T_0}. \tag{10}$$

Assuming all disturbances in terms of Fourier components, $\exp(i\mathbf{k}\cdot\mathbf{r} - i\omega t)$, where $\mathbf{k} = k_x \hat{x} + k_z \hat{z}$, we obtain the following algebraic equations:

$$\omega \rho_1 - \rho_0 (k_x v_{1x} + k_z v_{1z}) = 0, \tag{11}$$

$$\omega \rho_0 v_{1x} - k_x p_1 - \frac{B_0}{4\pi}(k_x B_{1z} - k_z B_{1x}) + \frac{\eta_0 i}{3}(k_x^2 v_{1x} - 2k_x k_z v_{1z}) = 0, \tag{12}$$

$$\omega \rho_0 v_{1y} + \frac{B_0}{4\pi} k_z B_{1y} = 0, \tag{13}$$

$$\omega \rho_0 v_{1z} - k_z p_1 + \frac{\eta_0 i}{3}(4k_z^2 v_{1z} - 2k_x k_z v_{1x}) = 0, \tag{14}$$

$$\omega B_{1x} + k_z B_0 v_{1x} = 0, \tag{15}$$

$$\omega B_{1y} + k_z B_0 v_{1y} = 0, \tag{16}$$

$$\omega B_{1z} - k_x B_0 v_{1x} = 0, \tag{17}$$

$$i\omega(p_1 - c_s^2 \rho_1) - (\gamma - 1)\kappa_\parallel k_z^2 T_1 = 0, \tag{18}$$

and

$$\frac{p_1}{p_0} - \frac{\rho_1}{\rho_0} - \frac{T_1}{T_0} = 0. \tag{19}$$

$\eta_0$ is coefficient of viscosity. We note here that Eqs. (11) - (19) are the same as Eqs. (24) - (32) of Kumar et al. (2006). The solution of the first set of equations (Eqs. (13) and (16)) in terms of variables $v_{1y}$ and $B_{1y}$ describes Alfvén waves in incompressible fluid. The solutions of the second set of equations in terms of variables $v_{1x}$, $v_{1z}$, $p_1$, $T_1$, $\rho_1$, $B_{1x}$ and $B_{1z}$ describe the damped magnetoacoustic waves in the x-z plane (cf., Field, 1965). Consequently, all other perturbation terms, namely $p_1$, $T_1$, $\rho_1$, $B_{1x}$, $B_{1z}$ must be eliminated in terms of $v_{1x}$ and $v_{1z}$, which are given as under.

From Eqs. (18) and (19), we get

$$p_1 = \left( \frac{c_s^2 + \frac{1}{\omega \rho_0} i(\gamma-1)\kappa_\| k_z^2 T_0}{1 + \frac{1}{\omega p_0} i(\gamma-1)\kappa_\| k_z^2 T_0} \right) \rho_1. \qquad (20)$$

When we substitute $B_{1x}$ and $B_{1z}$ from Eqs. (15) and (17) in Eqs. (12) and (14) we get:

$$\left( \omega^2 - v_A^2 k^2 + \frac{i}{3\rho_0}\eta_0 k_x^2 \omega \right) v_{1x} - \left( \frac{2i}{3\rho_0}\eta_0 k_x k_z \omega \right) v_{1z} = \frac{1}{\rho_0} k_x p_1 \omega, \qquad (21)$$

and

$$\left( \frac{2i}{3\rho_0}\eta_0 k_x k_z \right) v_{1x} - \left( \omega + \frac{4i}{3\rho_0}\eta_0 k_z^2 \right) v_{1z} = \frac{-1}{\rho_0} k_z p_1. \qquad (22)$$

When we put $p_1$ from Eq. (20) in Eqs. (21) and (22) and use Eq. (11), we get

$$\left( \omega^3 - v_A^2 k^2 \omega - c_s^2 k_x^2 \omega + \frac{i}{3\rho_0}\eta_0 k_x^2 \omega^2 + \frac{i}{p_0}(\gamma-1)\kappa_\| k_z^2 T_0 \omega^2 - \frac{1}{3p_0 \rho_0}\eta_0(\gamma-1)\kappa_\| k_x^2 k_z^2 T_0 \omega - \right.$$
$$\left. \frac{i}{p_0}(\gamma-1)\kappa_\| v_A^2 k^2 k_z^2 T_0 - \frac{i}{\rho_0}(\gamma-1)\kappa_\| k_x^2 k_z^2 T_0 \right) v_{1x} = \left( \frac{2i}{3\rho_0}\eta_0 k_x k_z \omega^2 - \right.$$
$$\left. \frac{2}{3p_0\rho_0}\eta_0(\gamma-1)\kappa_\| k_x k_z^3 T_0 \omega + c_s^2 k_x k_z \omega + \frac{i}{\rho_0}(\gamma-1)\kappa_\| k_x k_z^3 T_0 \right) v_{1z}$$

$$(23)$$

and

$$\left( c_s^2 k_x k_z \omega + \frac{2i}{3\rho_0} \eta_0 k_x k_z \omega^2 - \frac{2}{3p_0 \rho_0} \eta_0 (\gamma-1) \kappa_\| k_x k_z^3 T_0 \omega + \frac{i}{\rho_0}(\gamma-1)\kappa_\| k_x k_z^3 T_0 \right) v_{1x} = \left( \omega^3 - c_s^2 k_z^2 \omega + \frac{4i}{3\rho_0}\eta_0 k_z^2 \omega^2 + \frac{i}{p_0}(\gamma-1)\kappa_\| k_z^2 T_0 \omega^2 - \frac{4}{3p_0 \rho_0}\eta_0(\gamma-1)\kappa_\| k_z^4 T_0 \omega - \frac{i}{\rho_0}(\gamma-1)\kappa_\| k_z^4 T_0 \right) v_{1z}.$$

(24)

Eqs. (23) and (24) are two sets of algebraic equations in terms of two independent variables $v_{1x}$ and $v_{1z}$. Setting the determinant of the coefficients of these two equations equal to zero, we obtain the dispersion relation as:

$$\omega^6 + iA\omega^5 - B\omega^4 - iC\omega^3 + D\omega^2 + iE\omega - F = 0, \qquad (25)$$

Where $A = 2c_0 + c_1$,

$$B = (c_s^2 + v_A^2)k^2 + c_0(2c_1 + c_0),$$

$$C = c_2 + c_0(k^2(c_s^2 + 2v_A^2 + \frac{p_0}{\rho_0}) + c_0 c_1),$$

$$D = c_s^2 c_6 + c_0(c_3 + c_0 c_4),$$

$$E = c_0(c_0 c_5 + c_6(c_s^2 + \frac{p_0}{\rho_0})),$$

$$F = c_0^2 c_6 \frac{p_0}{\rho_0}.$$

and $\quad c_0 = \frac{1}{p_0}(\gamma-1)\kappa_\| k_z^2 T_0$,

$$c_1 = \frac{1}{3\rho_0}\eta_0(k_x^2 + 4k_z^2),$$

$$c_2 = \frac{1}{3\rho_0}\eta_0 k_z^2(4v_A^2 k^2 + 9c_s^2 k_x^2),$$

$$c_3 = \frac{1}{3\rho_0}\eta_0 k_z^2 (8v_A^2 k^2 + 9(c_s^2 + \frac{p_0}{\rho_0})k_x^2),$$

$$c_4 = (v_A^2 + \frac{p_0}{\rho_0})k^2,$$

$$c_5 = \frac{1}{3\rho_0}\eta_0 k_z^2 (4v_A^2 k^2 + 9\frac{p_0}{\rho_0} k_x^2),$$

$$c_6 = v_A^2 k^2 k_z^2.$$

## *2.1. Kumar et al.'s derivation*

Kumar et al. (2006) did not substitute the value of $p_1$ in the momentum equations. Instead, they obtained an additional equation from Eqs.(11) and (20). This resulted in three independent variables (erroneously defining the x-z plane in terms of three independent variables). Accordingly, they obtained three sets of algebraic equations, two from momentum equation and third from Eqs. (11) and (20), in terms of three independent variable $v_{1x}$, $v_{1z}$ and $p_1$, i.e.,

$$\left(\omega^2 - v_A^2 k^2 + \frac{i}{3\rho_0}\eta_0 k_x^2 \omega\right)v_{1x} - \left(\frac{2i}{3\rho_0}\eta_0 k_x k_z \omega\right)v_{1z} - \frac{1}{\rho_0}k_x \omega p_1 = 0 \ , \quad (26)$$

$$\left(\frac{2i}{3\rho_0}\eta_0 k_x k_z\right)v_{1x} - \left(\omega + \frac{4i}{3\rho_0}\eta_0 k_z^2\right)v_{1z} + \frac{1}{\rho_0}k_z p_1 = 0, \quad (27)$$

and

$$\left(c_s^2 k_x \rho_0 \omega + i(\gamma-1)\kappa_\| k_x k_z^2 T_0\right)v_{1x} + \left(c_s^2 k_z \rho_0 \omega + i(\gamma-1)\kappa_\| k_z^3 T_0\right)v_{1z}$$
$$- \left(\omega^2 + \frac{i}{p_0}(\gamma-1)\kappa_\| k_z^2 \omega T_0\right)p_1 = 0 \quad . \quad (28)$$

When we set the determinant of the coefficients of these three sets of algebraic equations equal to zero, we obtain a five-order dispersion relation of Kumar et al. (2006), i.e.,

$$\omega^5 + iA\omega^4 - B\omega^3 - iC\omega^2 + D\omega + iE = 0 \quad (29)$$

where $A = c_0 + c_1$, $B = (c_s^2 + v_A^2)k^2 + c_0 c_1,$

$$C = \left(c_0 c_2 + c_4 + c_s^2 c_3\right), \quad D = \left(c_0 c_4 + c_s^2 c_5 + \frac{p_0}{\rho_0} c_0 c_3\right), \quad E = \frac{p_0}{\rho_0} c_0 c_5.$$

$$c_0 = \frac{1}{p_0}(\gamma - 1)\kappa_\| k_z^2 T_0, \quad c_1 = \frac{1}{3\rho_0}\eta_0(k_x^2 + 4k_z^2), \quad c_2 = k^2 v_A^2 + \frac{p_0}{\rho_0} k^2,$$

$$c_3 = \frac{3}{\rho_0}\eta_0 k_x^2 k_z^2, \quad c_4 = \frac{4}{3\rho_0}\eta_0 k_z^2 k^2 v_A^2, \quad c_5 = v_A^2 k_z^2 k^2.$$

Here $c_s^2 = \frac{\gamma p_0}{\rho_0}$ cm²s⁻², $v_A^2 = \frac{B_0^2}{4\pi\rho_0}$ cm²s⁻², $\eta_0 = 10^{-16} T_0^{5/2}$ gm cm⁻¹s⁻¹ and $\kappa_\| = 10^{-6} T_0^{5/2}$ gm cm s⁻³ deg⁻¹ (Braginskii 1965; Porter et al. 1994).

Carbonell et al. (2004) have also derived a five-order dispersion relation, but for a different plasma configuration in which the direction of magnetic field is along x-axis. Moreover, they have not considered the effect of viscosity as a damping mechanism. It is to be noted here that in deriving the dispersion relation, Carbonell et al. (2004) have eliminated the perturbations $p_1, T_1, \rho_1, B_{1x}, B_{1z}$ in favour of $v_{1x}$ and $v_{1z}$ which resulted in two algebraic equations for the velocity perturbations. This is in agreement with our approach (cf., Eqs. (23) and (24)) but it is in disagreement with Kumar et al. (2006) approach.

## *2.2. Porter et al.'s derivation*

Our dispersion relation is in agreement (except some coefficients) with the results of Porter et al. (1994) but it is in disagreement with the previous results obtained by Kumar et al (2006). The wave heating terms $Q_{visc}$ and $Q_{th}$ are second order in energy equation. In order to have equilibrium state, Porter et al. (1994) replaced $Q_{rad}$ by $\lambda^2 Q_{rad}$, which is relevant only when we calculate energy damping rate by applying small damping approximation. However, amplitude damping rate $(\text{Im}(\omega))$ is calculated from the dispersion relation which is derived under the first order approximation. Consequently, the wave heating terms (second order terms) are neglected in the derivation of dispersion relation. This way, linearized forms and sets of algebraic equations will be the same as given in our derivation. The dispersion relation of Porter et al. (1994) supports thermal mode even in the absence of thermal conductivity about which we discuss further in the following section.

### 3. Results and discussion

Kumar et al. (2006) have derived a five-order dispersion relation by taking 3×3 determinant of the coefficients equal to zero. These coefficients in terms of three independent variables $v_{1x}, v_{1z}$, and $p_1$ appear in three algebraic equations as already noted. If we follow this approach, we do not get the inequality conditions $v_{1z} \gg v_{1x}$ (for slow mode waves) and $v_{1x} \gg v_{1z}$ (for fast mode waves) on which the weak damping approximation is valid (cf., Porter et al., 1994), due to an additional term, $p_1$. On the other hand, if we follow Porter et al. (1994) we get two sets of algebraic equations in terms of two independent variables $v_{1x}$ and $v_{1z}$. When we set 2×2 determinant of the coefficients equal to zero, we obtain a six-order dispersion relation.

The six-order dispersion relation has the dissipative terms of viscosity and thermal conductivity. Solution of this dispersion relation provides six roots namely, $\omega_{1r} - i\omega_{1i}$, $-\omega_{1r} - i\omega_{1i}$, $\omega_{2r} - i\omega_{2i}$, $-\omega_{2r} - i\omega_{2i}$, $\omega_{3r} - i\omega_{3i}$ and $-\omega_{4r} - i\omega_{4i}$, where $\omega_{3r}$ and $\omega_{4r}$ are negligibly small compared to $\omega_{1r}$ and $\omega_{2r}$. Thus two roots are purely imaginary which correspond to thermal mode and the other four roots are in the pair form. One pair corresponds to slow mode and the other pair to fast mode. If we consider the thermal conductivity only, we get six roots i.e., $\omega_{1r} - i\omega_{1i}$, $-\omega_{1r} - i\omega_{1i}$, $\omega_{2r} - i\omega_{2i}$, $-\omega_{2r} - i\omega_{2i}$, $\omega_{3r} - i\omega_{3i}$ and $-\omega_{4r} - i\omega_{4i}$. This means, we have slow mode, fast mode and thermal mode. When we consider the viscosity term only, we get four roots i.e., $\omega_{1r} - i\omega_{1i}$, $-\omega_{1r} - i\omega_{1i}$, $\omega_{2r} - i\omega_{2i}$ and $-\omega_{2r} - i\omega_{2i}$. This simply means that the thermal mode is excited only when thermal conductivity is present in the dispersion relation. It is to be further noted that the coefficients in the dispersion relation of Porter et al. (1994) do have the dissipative terms of viscosity, and thermal conductivity. Solution of this dispersion relation provides six roots namely $\omega_{1r} - i\omega_{1i}$, $-\omega_{1r} - i\omega_{1i}$, $\omega_{2r} - i\omega_{2i}$, $-\omega_{2r} - i\omega_{2i}$, $\omega_{3r} - i\omega_{3i}$ and $-\omega_{4r} - i\omega_{4i}$. Thus two roots are purely imaginary corresponding to thermal mode and the other four roots are in the pair form, one pair corresponds to slow mode and the other to fast mode. If we consider thermal conductivity only (i.e., $\eta_0 = 0$ but $\kappa_\parallel \neq 0$) we get five roots, having the form $\omega_{1r} - i\omega_{1i}$, $-\omega_{1r} - i\omega_{1i}$, $\omega_{2r} - i\omega_{2i}$, $-\omega_{2r} - i\omega_{2i}$ and $\omega_{3r} - i\omega_{3i}$. Obviously, one root is purely imaginary. Consequently, we get slow mode, fast mode and one root corresponding to thermal mode. If we consider viscosity only (i.e., $\eta_0 \neq 0$ but $\kappa_\parallel = 0$), we again get five roots, having the form $\omega_{1r} - i\omega_{1i}$, $-\omega_{1r} - i\omega_{1i}$, $\omega_{2r} - i\omega_{2i}$, $-\omega_{2r} - i\omega_{2i}$ and $\omega_{3r} - i\omega_{3i}$, which will result in slow mode, fast mode and thermal mode. It is to be further noted that the dispersion relation of Porter et al. (1994) contains thermal mode even in the absence of thermal conductivity. This is in conflict with the results of De Moortel and Hood (2003) that the thermal modes can exist in the presence of thermal conductivity only.

In conclusion, we have shown that the actual dispersion relation for MHD waves in a homogeneous medium must be six-order. Our finding is in agreement (except some coefficients) with the results of Porter et al. (1994) but it is in disagreement with the previous results obtained by Kumar et al. (2006).

**Acknowledgements**

V.S. Pandey acknowledges the CSIR, New Delhi for the award of a senior research fellowship. We thank the referee for helpful comments.